\documentclass[letter]{aa}
\usepackage{graphicx}
\usepackage{txfonts}
\usepackage{natbib}
\usepackage{verbatim}
\usepackage{longtable}
\usepackage{lscape}
\usepackage{afterpage}
\usepackage{multirow}
\bibpunct{(}{)}{;}{a}{}{,}
\defcitealias{FM07}{FM07}

\begin{document}

\title{The role of the charge state of PAHs in ultraviolet extinction}

\author{C. Cecchi\textendash Pestellini\inst{1}
	\and
	G. Malloci\inst{1}
	\and
	G. Mulas\inst{1}
	\and
	C. Joblin\inst{2}
	\and
	D.~A.~Williams\inst{3}
}

\institute{
Istituto Nazionale di Astrofisica~\textendash~Osservatorio Astronomico di Cagliari, 
Strada n.54, Loc. Poggio dei Pini, I\textendash09012 Capoterra (CA), Italy
\textendash\email{ccp@ca.astro.it}
\and
Centre d'Etude Spatiale des Rayonnements, Universit\'e de Toulouse
\textendash CNRS, Observatoire Midi-Pyr\'en\'ees, 9 Avenue du Colonel Roche,
31028 Toulouse cedex 04, France
\and
University College London, Department of Physics and Astronomy,
Gower Street, London WC1E 6BT, UK
}
\date{Received 21 April 2008; Accepted 9 June 2008}

\abstract
{}
{We explore the relation between the charge state of polycyclic aromatic 
hydrocarbons (PAHs) and the extinction curve morphology.}
{We fit extinction curves with a dust model including core\textendash mantle 
spherical particles of mixed chemical composition (silicate core, $sp^2$ and 
$sp^3$ carbonaceous layers), and an additional molecular component. We use 
exact methods to calculate the extinction due to classical particles  
and accurate computed absorption spectra of PAHs in different charge states,
for the contribution due to the molecular component, along five different 
lines of sight.}
{A combination of classical dust particles and mixtures of real PAHs
satisfactorily matches the observed interstellar extinction curves. Variations 
of the spectral properties of PAHs in different charge states
produce changes consistent with the varying relative strengths of the bump 
and non\textendash linear far\textendash UV rise.}
{}


\authorrunning{Cecchi\textendash Pestellini et al.}
\titlerunning{The role of charge state of PAHs in UV extinction}

\maketitle

\section{Introduction}
The nature of interstellar dust has implications for interstellar extinction, 
scattering of starlight, interstellar chemistry, the heating of diffuse 
clouds, the deposition of interstellar ices, and the dynamics of star 
formation.

Models of interstellar dust take into account a wide range of observational
information that constrains the composition and size distribution \citep{D03}. 
This includes the pattern of elemental depletions, the variety of observed 
interstellar extinction curves, the distribution of the linear 
polarisation of starlight, the properties of scattered light, absorption and 
emission features, and broadband emission in the visible and infrared regions.
All models involve a significant fraction of dust in the form of solid carbon,
some of which may be in the form of crystalline graphite \citep{T97}. Other 
forms of carbon have been given various names (amorphous, 
diamond-like, glassy and hydrogenated amorphous carbon, quasi-carbonaceous 
condensate, soot, yellow stuff\ldots) yet these all seem
manifestations of carbon in which particular valence long\textendash 
range structures and physical/chemical properties (e.g., hydrogen content) 
are emphasised. 

The strongest interstellar extinction feature is the broad absorption 
bump at 217.5~nm discovered by \citet{S65}, whose carrier is as yet unknown. 
It has been attributed to a plasmon resonance associated with 
$\pi \to \pi^\star$ transitions of electrons in 
graphite \citep{SD65}, despite some difficulties with this 
interpretation \citep[e.~g.][]{DM93}. Others have tried to associate the bump 
with amorphous carbon \citep{M98}, although the peak and the width of the 
feature do not seem to be simultaneously well reproduced. It has also 
been suggested that defect sites at the surface of the silicate 
material could carry the feature \citep{SD87}. The intensity of the 
bump is known to be unrelated to the far\textendash UV rise of the 
interstellar extinction curve \citep[ISEC,][]{gre83}, 
while its FWHM shows an extremely loose relation to the strength of the 
nonlinear far\textendash UV curvature \citep[][ henceforth 
\citetalias{FM07}]{FM07}, 
although with an intrinsic scatter much greater than the observational errors. 

Leaving aside specialised hypotheses, almost all
the proposed carriers of the bump seem to require some 
form of aromatic carbon, either as size\textendash restricted graphite pieces 
\citep{D85} or as single or stacked Polycyclic Aromatic Hydrocarbons 
(PAHs) \citep{DS98,LD01b,D06}. The possible quantitative relation between the 
bump and PAHs was suggested by \citet{J92}, who compared
laboratory spectra of mixtures of neutral PAHs and the mean galactic 
ISEC, showing them to be compatible in spectral shape. Subsequent papers
\citep{mal04,mal05,mal07a} compared theoretical spectra
of mixtures of PAHs in different charge states.

Recent models have a significant fraction of carbon in PAH 
molecules or clusters, described using approximate average properties 
\citep{WD01,LD01,zub04}. In this work we incorporate PAHs with
their real extinction properties in the evolutionary model of interstellar 
dust proposed by \citet{JDW90}. In the latter model, 
carbonaceous material is in the form of mantles on the surfaces of silicate 
cores, forming through deposition processes conceptually similar to 
the grain core and icy mantle model of dust grains in dark clouds. 
The core\textendash mantle structure and chemical composition of dust particles result 
from the history of the environmental conditions they experienced, and 
responded to, so that the observed extinction along any line of sight is an 
integration of components at different stages of their evolution \citep{CPW98}.
Different core\textendash mantle models propose different accretion schemes for the 
carbonaceous mantles, either through the cyclic processing of ices \citep[as 
in][]{LG97} or through direct condensation \citep[as here, following][]{JDW90}.
Within such a framework, PAHs must also respond to the same environmental 
conditions, with chemical changes possibly related to observable effects.  

We investigate the relation between the optical properties of PAHs in 
different ionisation states with the ISEC. We take advantage of the 
recent availability of state\textendash of\textendash the\textendash art 
absorption spectra of PAHs in several charge states \citep{mal07b}, and 
the extinction cross\textendash sections of classical 
core\textendash mantle dust grains computed through an exact approach 
based on the multipole expansion of the electromagnetic fields \citep{BDS07}. 
Synthetic extinction patterns are consistent with the observed properties of 
the carriers of the bump and of the non\textendash linear far\textendash UV 
rise. We propose a quantitative interpretation of the relative strengths of 
these features in terms of varying charge states of PAHs.

\section{The model}\label{model}

While considerable uncertainty remains about the physical nature of 
interstellar dust, in particular its shape and morphology 
\citep[e.~g.][]{Wr87}, spectroscopic evidence and the depletion data
show that silicates and carbons are important components of it. Silicates are 
predominantly amorphous but with a crystalline component \citep{LD01}. 
Carbons include both graphitic, $sp^2$, and polymeric, $sp^3$, valences 
\citep[e.~g.][]{JDW90}. They are mostly amorphous, 
with some evidence of the presence of crystalline graphite.

It is unclear whether the silicates and carbons form distinct populations of 
dust grains or are mixed within the same particles, since both hypotheses 
fit the ISEC. \citet{chi06} showed that the excess polarisation in the 
3.4~$\mu$m aliphatic C\textemdash H stretch observed towards the 
galactic centre is much less than that of the 9.7~$\mu$m silicate feature; 
they interpreted this to rule out that the two features could
arise from the same population of dust particles. However, this conclusion 
is not necessarily valid if the two features arise from 
\emph{different parts} of the same dust grain population, e.g., cores 
and mantles \citep{LG02}.

We use the extinction model of \citet{CPW98} and \citet{IAP}, i.~e. a 
core~+~multiple mantle model, with dust grains composed of a silicate 
core and two carbon mantles. The deeper carbonaceous layer represents 
"older", UV processed $sp^2$ material, and the surface layer "freshly deposited"
$sp^3$ material. For the former we adopted the BE optical constants
of \citet{rou91,rou93}, which are the most representative for $sp^2$ amorphous
carbon \citep[see][for a more detailed discussion]{ccp97}; for $sp^3$ we 
used the optical constants for polymeric carbon of \citet{ash91}.

Following \citet{CPW98} and \citet{IAP}, the 
silicate cores are a population of small and large spherical grains 
with a power law size distribution $a^{-q}$, $a$ being the core radius.
Silicate grains that are accumulating carbon 
mantles will all have the same thickness of mantle \citep{W02}. This component 
is then completely determined by mantle thickness 
and $sp^2$ fraction. The contribution of PAHs to extinction is represented 
in the \citet{IAP} model by
two Lorentz profiles, one for the $\pi\to\pi^\star$ transitions, 
accounting for the bump, the second for the $\sigma\to\sigma^\star$ 
plasmon resonance, whose low\textendash energy tail contributes to the 
far\textendash UV rise. Such a model reproduces the observed average ISEC with 
good accuracy, requiring an amount of carbon within the budget available in the
interstellar medium \citep[for more details see][]{IAP}. 

For a physical interpretation of the molecular component we exploit the  
database of spectral properties of PAHs computed by \citet{mal07b}, including 
50 molecules in the size range 10\textendash66 C atoms, 
in the charge states 0, $\pm$1, and +2. For details see \citet{mal07b}.

\begin{figure}[t!]
\includegraphics{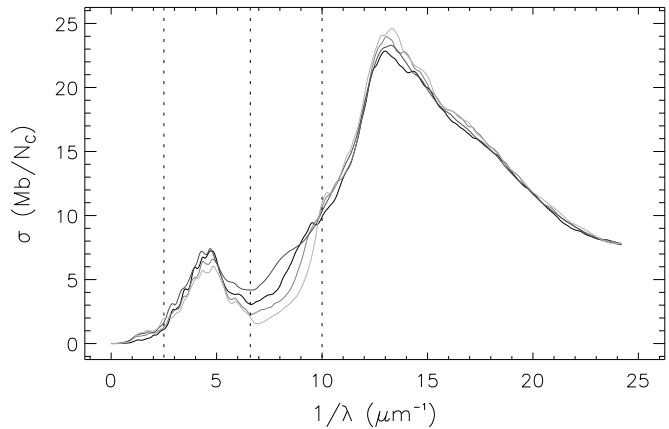}
\caption{Weighted sums of cross\textendash sections of PAHs in 
different charge states, in Mb ($10^{-18}$~cm$^2$) per number of 
carbon atoms N$_{\rm C}$. The curves correspond to neutrals (black), anions 
(dark gray), cations (gray), and dications (light gray). The vertical 
dashed lines show the intervals corresponding to the bump and the 
non\textendash linear far\textendash UV rise (see Table~\ref{intweighted}).
\label{weighted}}
\end{figure}

Figure~\ref{weighted} displays the weighted sums of the spectra in our sample,
weighted with the inverse of the number of C atoms of each molecule. 
The two resonances corresponding to the bump and the far\textendash UV rise 
overlap near $6.5~\mu\mathrm{m}^{-1}$, and their 
relative intensity varies with charge state, as does the depth of the 
minimum between them. This is systematic for each molecule and depends 
roughly on charge state and molecular size, yielding a 
2\textendash dimensional manifold. We show average values for clarity. 
In Table~\ref{intweighted} we list, for each sum, the ratio between 
the integrated intensities of the bump and of the far\textendash UV rise, 
and the cross\textendash section per C atom at the minimum between 
them. This systematic behaviour hints that PAH charge state is linked 
to morphology properties of the observed ISECs, motivating this work: 
\citetalias{FM07} showed that the non\textendash linear far\textendash UV rise fluctuates widely 
in both strength and slope; our calculations suggest that their parametrisation
of the ISECs is linked to the relative strength and position of the $\pi\to\pi^\star$ 
and $\sigma\to\sigma^\star$ resonances.

\begin{table}[b!]
\begin{center}
\caption{\label{intweighted} 
For each sum in Fig.~\ref{weighted}, first row: 
ratio between its integrals in the bump
(f$_\mathrm{bump}$, 2.5 to 6.6~$\mu$m$^{-1}$), and the 
far\textendash UV rise (f$_\mathrm{rise}$, 6.6 to 10~$\mu$m$^{-1}$)
regions; 
second row: cross\textendash section (in Mb per C atom) at the minimum 
between the $\pi\to\pi^\star$ and $\sigma\to\sigma^\star$ resonances.}
\begin{tabular}{ccccc}
\hline \hline
\noalign{\smallskip}
& Anions & Neutrals & Cations & Dications\\
\noalign{\smallskip}
\hline
\noalign{\smallskip}
$\displaystyle \frac{f_\mathrm{bump}}{f_\mathrm{rise}}$ & 0.89 & 0.92 & 1.01 
& 1.23 \\
\noalign{\smallskip}
$\sigma_\mathrm{min}$ & 4.18 & 3.06 & 2.27 & 1.55 \\
\noalign{\smallskip}
\hline
\end{tabular}
\end{center}
\end{table}

\section{Results} \label{results}
We modelled five extinction curves taken from the 
sample of \citetalias{FM07}, with $R_V$ ranging from 2.33 to 5.05, 
$R_V$ being the ratio of total to selective extinction. The fitting 
procedure provides two descriptions for the classical component. For those 
ISECs with $R_V \lesssim 4$, the parameter set exploited by \citet{IAP} is 
appropriate, while for larger $R_V$ the derived grain distribution is
significantly different: the small particle component is missing, large 
silicate cores have radii larger than 100~nm, and the index $q$ of the 
power\textendash law size distribution has the standard value of 3.5. 
This could indicate depletion of small grains due to coagulation 
processes \citep{ste03}. This would produce a tail of very large particles
in the size distribution, with a negligible effect on the modelled 
extinction, to which our fit is therefore insensitive. The abundances 
reported in Table~\ref{table} for the classical dust component do not include 
the possible contribution of this tail of large particles, and 
thus should be considered with some caution as lower limits.
For the whole set of ISECs the total mantle thickness is 1.25~nm, while the 
$sp^2$ fraction is almost constant between $0.93$ and $0.98$. 
This large $sp^2$ fraction is consistent with the short lifetime of the CH 
bond inferred by laboratory experiments on interstellar grain analogues 
\citep{M01}. The presence of a small $sp^3$ fraction resulting
from the fit, negligible for the extinction in the ISECs considered, 
is consistent with the physical evolutionary model. The full set of model 
parameters obtained from the fit are reported in Table~\ref{table}.
\begin{table*}
\begin{center}
\caption{In the first part from the top,
A$_\mathrm{V}$, R$_\mathrm{V}$, $c_3$ 
(intensity of the bump), and $c_4$ (intensity of the non\textendash linear 
far\textendash UV rise) from \citet{FM07}. 
Second part: best\textendash fitting parameters of the 
core + multiple\textendash mantle classical particles model. Their
meanings are explained in Sect.~\ref{model}. We also list some physical
quantities of astronomical interest derived from them, namely the abundances
of Si and C locked in this component, relative to total H.
Third part: total column density N$_\mathrm{C}^\mathrm{PAH}$ of C in 
PAHs, its abundance relative to total H, its fraction in each 
charge state, and the average charge (in e$^-$ charges) per C atom, as 
obtained from the fit. The reported
standard deviations include only statistical errors, not systematic ones due
to the assumptions in the fit. All abundances are computed assuming 
$\mathrm{N}_\mathrm{H} \simeq 5.9 \times 10^{21} 
\mathrm{A}_\mathrm{V}/\mathrm{R}_\mathrm{V}$, with
 $\mathrm{N}_\mathrm{H}$ in cm$^{-2}$ and $\mathrm{A}_\mathrm{V}$ in mag
\citep{W02}. \label{table}}
\begin{tabular}{cccccc}
\hline \hline
\noalign{\smallskip}
Parameter & HD& HD & NGC 
& HD & HD\\
& 220057 & 197512 & 457~Pech13 
&147889 & 93222\\
\noalign{\smallskip}
\hline
\noalign{\smallskip}
A$_\mathrm{V}$ & 0.56 & 0.71 & 1.58 
& 4.44 & 2.17 \\
R$_\mathrm{V}$ & 2.333 & 2.433 & 3.106 
& 4.193 & 5.046 \\
$c_3$ & 2.665 & 4.136 & 2.906 
& 4.228 & 1.942\\
$c_4$ & 0.374 & 0.500 & 0.256 
& 0.525 & 0.250\\
\noalign{\smallskip}
\hline
\noalign{\smallskip}
$w$ (nm) & 1.25 & 1.25 & 1.25 & 1.25 & 1.25 \\
\noalign{\smallskip}
$\displaystyle {w_\mathrm{sp^2}}/{w}$ & 0.93 & 0.94 & 0.96 & 0.96 & 0.98 \\
\noalign{\smallskip}
$\displaystyle \left(a_\mathrm{min} - a_\mathrm{max} \right)_\mathrm{small}$ 
(nm) & 5\textendash12 & 5\textendash12 & 5\textendash12 & \textemdash& 
\textemdash\\
\noalign{\smallskip}
$\displaystyle \left(a_\mathrm{min} - a_\mathrm{max} \right)_\mathrm{large}$
(nm) & 42\textendash250 & 42\textendash250 & 42\textendash250 & 
110\textendash250 & 110\textendash250 \\
\noalign{\smallskip}
MRN index & 3.531 & 3.531 & 3.531 & 3.5 & 3.5 \\
\noalign{\smallskip}
N$_\mathrm{C}^\mathrm{dust}$/N$_\mathrm{H}$ (ppM)$^\diamond$ & 43 & 37 & 57 & 14 & 12 \\
\noalign{\smallskip}
N$_\mathrm{Si}^\mathrm{dust}$/N$_\mathrm{H}$ (ppM)$^\diamond$ & 36 & 31 & 47 & 11 & 10 \\
\noalign{\smallskip}
\hline
\noalign{\smallskip}
N$_\mathrm{C}^\mathrm{PAH}$ ($10^{16}$ cm$^{-2}$) & $5.92\pm0.48$ & 
$18.5\pm1.0$ & $16.4\pm0.5$ & $108\pm4$ & $42.7\pm 1.9$\\
\noalign{\smallskip}
N$_\mathrm{C}^\mathrm{PAH}$/N$_\mathrm{H}$ (ppM) & 42 & 107 & 55 & 173 & 168 \\
\noalign{\smallskip}
N$_\mathrm{C}^\mathrm{PAH^{-1}}$/N$_\mathrm{C}^\mathrm{PAH}$ 
& $0.62\pm0.20$ & $0.76\pm0.07$ & $0.75\pm0.03$ 
& $0.75\pm0.10$ & $0.65\pm0.05$\\
\noalign{\smallskip}
N$_\mathrm{C}^\mathrm{PAH^0}$/N$_\mathrm{C}^\mathrm{PAH}$ 
& $0.34\pm0.18$ & $0.15\pm0.05$ & $0.25\pm0.03$ 
&$0.20\pm0.09$ & $0.16\pm0.05$\\ 
\noalign{\smallskip}
N$_\mathrm{C}^\mathrm{PAH^{+1}}$/N$_\mathrm{C}^\mathrm{PAH}$ 
& $0.03\pm0.06$ & $0.06\pm0.03$ & $0.00\pm0.01$
& $0.03\pm0.02$ & $0.16\pm0.04$\\
\noalign{\smallskip}
N$_\mathrm{C}^\mathrm{PAH^{+2}}$/N$_\mathrm{C}^\mathrm{PAH}$ 
& $0.01\pm0.02$ & $0.04\pm0.03$ & $0.01\pm0.01$
& $0.02\pm0.03$ & $0.03\pm0.02$\\
\noalign{\smallskip}
Charge/N$_\mathrm{C}^\mathrm{PAH}$
& $-0.025\pm0.011$ & $-0.043\pm0.005$ & $-0.043\pm0.002$
& $-0.041\pm0.008$ & $-0.037\pm0.004$\\
\noalign{\smallskip}
\hline
\noalign{\smallskip}
N$_\mathrm{C}^\mathrm{total}$/N$_\mathrm{H}$ (ppM)$^\diamond$ & 85 & 144 & 112 
& 187 & 180 \\
\noalign{\smallskip}
\hline
\end{tabular}
\end{center}
$^\diamond$Lower limits, since the possible presence of large dust 
is not included here, see text in Sect.~\ref{results}.
\end{table*}

\begin{figure}
\includegraphics{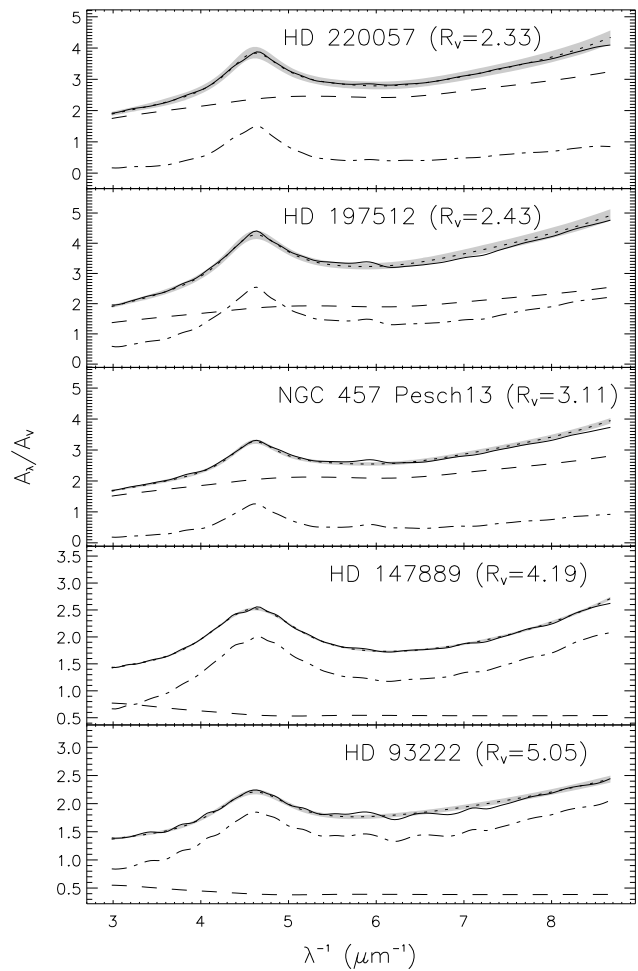}
\caption{
Comparison between the synthetic extinction curves, with best fitting 
parameters (continuous line) and the observed extinction curves 
\citepalias[dotted line][]{FM07}, with the 1$\sigma$ region in 
gray shade. The dashed and dash\textendash dotted lines are the separate 
contributions respectively of the core\textendash mantle grains and of 
the PAH mixtures.
\label{ccpfit}}
\end{figure}

To isolate the effect of the molecular component, we subtract from each 
modelled ISEC the extinction due to the dust. We then fit the residual with a 
linear combination of computed cross\textendash sections of PAHs. 
We adopt as free parameters the number of C atoms in each species, for a 
total of 200, with the physical constraint that they must all be positive. 
Finally, we sum the dust contribution with the best\textendash fitting 
combination of PAH spectra to obtain the synthetic 
extinction curve to be compared with observations. As shown in 
Fig.~\ref{ccpfit}, the fit to the observations is quantitatively 
excellent in all cases, in the whole spectral range: the synthetic
curves are almost always within the 1$\sigma$ region, and never 
further than 2$\sigma$ from the observations.

From a geometric point of view, the $\chi^2$ hypersurface of 
this fit is a 200\textendash dimensional paraboloid. Its vertex is the formal 
unconstrained solution, its curvature tensor is the inverse of the 
covariance matrix. The paraboloid is not ``bowl shaped'', but more like a 
trough, narrow along a small number of directions $(\lesssim10)$ and shallow 
(or even almost flat) in the others. This means that some linear 
combinations of the free parameters are strongly constrained by the fit, 
while others are almost free.

The most strongly constrained combination of parameters (i.~e. the
direction of largest principal curvature of the paraboloid) is
proportional to the total number of C atoms N$_\mathrm{C}^\mathrm{PAH}$ in all 
species, while the combinations that are almost free (i.~e. the
directions in which the paraboloid is flat) shuffle C atoms among
species without changing the global charge of the mixture. This is 
consistent with the interpretation of the bump and far\textendash UV 
non\textendash linear rise in terms of global properties of the whole class of PAHs, 
rather than a feature associated with some individual molecule.
To find out to what extent, quantitatively, charge, as a global 
property, is well\textendash determined, we repeated the fit 20 times for 
each ISEC, each time perturbing it by white noise, proportionally to the 
errors reported by \citetalias{FM07}. The total N$_\mathrm{C}^\mathrm{PAH}$, its 
fraction in each charge state and average charge were computed each time, 
yielding synthetic statistics stemming from the errors in the ISECs. Results 
are reported in Table~\ref{table} of the Appendix. The statistical 
standard deviations obtained show that these collective properties of PAHs 
are indeed well\textendash defined by the fit, within the given assumptions. 
In Fig.~\ref{ccpfit} we also show the extinctions due to 
core\textendash mantle grains. The curves present similar trends:
classical dust grains provide an (approximately) linear baseline over 
which the contribution due to the molecular component lies. This
is consistent with previous models incorporating PAHs 
\citep{LD01,WD01,zub04}. Since the contribution of the classical component 
is linear in the spectral region of the non\textendash linear 
far\textendash UV rise, this feature is entirely due to the molecular 
component.

\section{Conclusions and future work}
\begin{figure}
\includegraphics{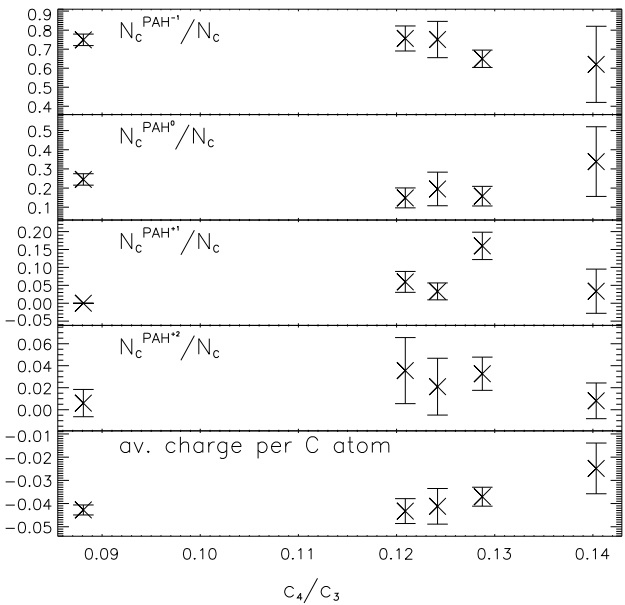}
\caption{Comparison of the ratio between the $c_4$ and $c_3$ parameters 
of \citetalias{FM07} and the physical parameters derived from our fitting
procedure for a sample of ISECs.
}
\label{chargecorrfm}
\end{figure}

Figure~\ref{chargecorrfm} compares the ionisation fractions in 
Table~\ref{table} with the ratio between the $c_4$ (intensity of the 
non\textendash linear far\textendash UV rise) and $c_3$ (intensity of 
the bump) parameters of \citetalias{FM07}. It is consistent 
with our \emph{ansatz} that the charge state of PAHs is connected to the 
observed variations in the relative intensity of the two features, even 
if it is probably not the sole factor determining it. A detailed, systematic
study will be needed to disentangle the relations between other specific 
properties of the PAH population and specific parameters of the ISECs.

This result is not strongly dependent on the model used to subtract the dust 
contribution to extinction: changing the parameters of the classical dust 
component has effects that are orthogonal to those brought about by the 
molecular component. For example, reasonable changes in the size cut\textendash offs and 
in the index of the power\textendash law size distribution only slightly modify the 
linear baseline on top of which the contribution of the molecular component 
lies (cf. Fig.~\ref{ccpfit}). 

The validity of the physical interpretation we propose relies on the 
two assumptions that (i) the PAH spectra we used may represent 
the real interstellar molecular component, and (ii) the 
cross\textendash sections do not change in a major way between 
free\textendash flying and clustered PAHs.

We include only fully hydrogenated, unsubstituted PAHs. 
\citet{D06} proposed completely dehydrogenated coronene 
as the carrier of the bump, emphasising that strongly dehydrogenated PAHs 
should display a plasmon resonance near the bump position. Hydrogenation is an 
additional degree of freedom that can respond to environmental conditions and
possibly play a role, together with charge state, in describing the variations 
observed in the ISEC. Since very little is known about the UV spectra 
of dehydrogenated and substituted PAHs, more experimental and 
theoretical work is needed before this can be addressed.

The PAH spectra we used for the fit are known to slightly
underestimate the energy of the $\pi\to\pi^\star$ transitions,  
similarly across the whole sample \citep{mal04}. 
This, together with the limited number of species used to 
represent the whole class of interstellar PAHs, produces a systematic error 
in the fit. This is more important for ISECs with a narrow bump, as this 
reduces the number of matching molecules and increases the effect of the
misplacement of the $\pi\to\pi^\star$ transitions mentioned above.
Using only half of the molecules at random, the difference between the 
best fit and the ISECs doubles, and absolute ionisation fractions 
change by up to about twice the statistical errors displayed in 
Table~\ref{table}. These absolute values should therefore be 
considered with caution, even if some models indeed 
predict a significant population of PAH anions \citep{tie05} in
regions including the diffuse galactic ISM.
Still, total N$_\mathrm{C}^\mathrm{PAH}$ and the \emph{trends} 
in Fig.~\ref{chargecorrfm} are instead almost unaffected by variations
in the PAH sample used for the fit, since 
they are due to \emph{systematic} variations of the PAH spectra.
Reducing the number of species \emph{at random}
worsens the quality of the fits (i.e. the $\chi^2$) but still reproduces the 
variations in the ISECs considered; on the other hand, excluding 
\emph{charge states} (e.g. considering only neutrals and cations) 
has a much more dramatic effect, i.e. it becomes impossible to satisfactorily
match the observations.

As to point (ii), we assumed PAHs to be in free molecular form,
whereas they could also be clustered \citep{DS98,rap06}. Clusters of neutral 
PAHs are weakly bound systems in which the electron densities of 
individual units (and thus their absorption spectra) are not strongly 
perturbed. The effect of clustering may be more important in charged 
systems, but since this is poorly known it is impossible, at present, 
to consider it in more detail.
Our estimation is that both clustering and a larger chemical 
variety would smooth out fine structure in the resonances, 
approaching, in the limit of large clusters, bulk optical properties. 
This would bring the absorption spectra of such improved mixtures of PAHs 
towards better agreement with observations. This will need to be confirmed 
by future calculations.

Mixtures of PAHs can accurately account for the UV bump and non\textendash linear far\textendash UV 
rise in extinction, and the PAH charge state is linked to the relative 
intensity of the two features. This provides a physical explanation and 
quantitative relationship of the 2\textendash dimensional 
variations in ISECs described by \citetalias{FM07}.
In this framework, it is therefore not surprising that 
the bump and the non\textendash linear far\textendash UV 
rise appear to be unrelated from an observational point of view:
this reflects the variations of the spectral properties
of PAHs in different charge states.

\begin{acknowledgements}
C.~C.-P., G.~M., G.~M. acknowledge financial support by MIUR under project 
CyberSar, call 1575/2004 of PON 2000-2006. 
G.~Malloci acknowledges financial support by Regione Autonoma della Sardegna. 
Part of the calculations used here have been performed using CINECA 
supercomputing resources.
We thank the anonymous referee for her/his help in improving the manuscript.
\end{acknowledgements}

\end{document}